%% file: main_final.tex
\documentclass[lettersize,journal]{IEEEtran}
\usepackage{algorithm,algorithmic,amsmath,amssymb,amsthm,bbm,cite,color,comment,graphicx,microtype,setspace,url,caption}
\usepackage[USenglish]{babel}
\usepackage{hyperref}
\usepackage[utf8]{inputenc}
\usepackage[T1]{fontenc}
\usepackage[font=small]{caption}
\usepackage{lipsum}

\usepackage{enumitem}
\setitemize{itemsep=0mm,topsep=0mm,parsep=0pt,partopsep=0pt}
\usepackage{stfloats}
\usepackage{tikz}
\usetikzlibrary{arrows,backgrounds,calc,intersections,decorations.pathreplacing,positioning,shapes}
\makeatletter
\newcommand\subparagraph{%
  \@startsection{subparagraph}{5}
  {\parindent}
  {3.25ex \@plus 1ex \@minus .2ex}
  {-1em}
 {\normalfont\normalsize\bfseries}}
\makeatother
\usepackage{titlesec}
\let\subparagraph\relax

\usepackage{titlesec}

\usepackage{tikz}
\usetikzlibrary{arrows, arrows.meta, backgrounds, calc, intersections, decorations.markings, decorations.pathreplacing, positioning, shapes}

\usepackage{pgfplots}
\usepgfplotslibrary{groupplots}
\pgfplotsset{compat=1.17}

\usepackage[font=small]{caption}

\input{./Definitions.tex}

\newcommand{\bs}{\textnormal{\tiny{AP}}}
\newcommand{\dl}{\textnormal{\tiny{DL}}}
\newcommand{\mse}{\mathrm{MSE}}
\newcommand{\gmse}{\mathrm{G\textnormal{-}MSE}}
\newcommand{\ue}{\textnormal{\tiny{UE}}}
\newcommand{\ul}{\textnormal{\tiny{UL}}}
\newcommand{\dlgrp}{\textnormal{\tiny{DL-grp}}}
\newcommand{\ulA}{\textnormal{\tiny{UL-1}}}
\newcommand{\ulB}{\textnormal{\tiny{UL-2}}}
\newcommand{\tg}{\textnormal{\tiny{G}}}

\title{Combined DL-UL Distributed Beamforming Design for Cell-Free Massive MIMO}
\author{
Bikshapathi Gouda, Antti Arvola, Italo Atzeni, and Antti Tölli
\thanks{The authors are with the Centre for Wireless Communications, University of Oulu, Finland (e-mail: \{bikshapathi.gouda, antti.arvola, italo.atzeni, antti.tolli\}@oulu.fi). This work was supported by the Research Council of Finland (336449 Profi6, 343586 CAMAIDE, 346208 6G~Flagship, and 348396 HIGH-6G) and by the European Commission (101095759 Hexa-X-II).}}

\begin{document}

\maketitle

\begin{abstract}
We consider a cell-free massive multiple-input multiple-output system with multi-antenna access points (APs) and user equipments (UEs), where the UEs can be served in both the downlink (DL) and uplink (UL) within a resource block. We tackle the combined optimization of the DL precoders and combiners at the APs and DL UEs, respectively, together with the UL combiners and precoders at the APs and UL UEs, respectively. To this end, we propose distributed beamforming designs enabled by iterative bi-directional training (IBT) and based on the minimum mean squared error criterion. To reduce the IBT overhead and thus enhance the effective DL and UL rates, we carry out the distributed beamforming design by assuming that all the UEs are served solely in the DL and then utilize the obtained beamformers for the DL and UL data transmissions after proper scaling. Numerical results show the superiority of the proposed combined DL-UL distributed beamforming design over separate DL and UL designs, especially with short resource~blocks.
\end{abstract}

\vspace{-5mm}


\vspace{-4mm}

\section{Introduction} \label{sec:INTRO}

In cell-free massive multiple-input multiple-output (MIMO) systems, multiple access points (APs) are connected to a central unit (CU) and jointly serve all the user equipments (UEs), eliminating the inter-cell interference and providing a uniform service across the network~\cite{Ngo17,Dem21}. In this regard, network-wide beamforming designs based on mean squared error (MSE) minimization or zero forcing, which require global channel state information (CSI) of all the APs at the CU, outperform local beamforming designs based on maximum ratio transmission/combining, which require only local CSI at each AP~\cite{Nay17, Dem21}. However, optimizing the network-wide beamformers at the CU calls for extensive CSI exchange via backhaul links, which poses severe challenges in terms of delays, backhaul bandwidth, and scalability of the network.

In coordinated cellular systems with time-division duplexing (TDD), the beamformers for the multi-antenna APs and UEs can be locally designed using iterative bi-directional training (IBT) with pilot-aided uplink (UL) and downlink (DL) channel estimation~\cite{Tol19}. Remarkably, IBT applied to cell-free massive MIMO systems enables fully distributed beamforming design at each AP by incorporating effective CSI from the other APs via over-the-air (OTA) signaling~\cite{Atz21,Atz20}. Typically, the beamforming design for UEs served in either the DL or UL requires separate IBT procedures. However, it was shown in~\cite{Gou21a} that using a common IBT procedure for the DL and UL beamforming designs may reduce the IBT overhead and thus enhance the effective DL and UL rates. Nonetheless, \cite{Gou21a} was limited to a centralized design (i.e., at the CU) and assumed all the UEs to be served in both the DL and UL.

In this paper, we consider a cell-free massive MIMO system with multi-antenna APs and UEs, where the UEs can be served in both the DL and UL within a resource block. We optimize the DL precoders and combiners at the APs and DL UEs, respectively, together with the UL combiners and precoders at the APs and UL UEs, respectively. In contrast to prior works, we consider a general system model with partially overlapping DL and UL UEs, and propose fully distributed beamforming designs enabled by IBT. To reduce the IBT overhead, we carry out the distributed beamforming design by assuming that all the UEs are served solely in the DL. Then: for the DL data transmission, the DL precoders for the UEs served only in the UL are discarded and each AP's transmit power is redistributed among the precoders for the DL UEs; for the UL data transmission, the DL combiners at the UL UEs are utilized as UL precoders after proper scaling to satisfy the per-UE transmit power constraints. To further reduce the number of precoders at each AP (and thus the IBT overhead), UEs served only in the DL can be paired with UEs served only in the UL, where each pair is assigned a common DL multicast precoder during the IBT. Numerical results demonstrate the superiority of the proposed combined DL-UL distributed beamforming designs over separate DL and UL designs, especially with short resource~blocks.

\section{System Model} \label{sec:SM}

\begin{figure}[t!]
\begin{center}
\begin{tikzpicture}

\pgfdeclareimage[height=8mm]{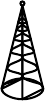}{figures/BS}
\pgfdeclareimage[height=6mm]{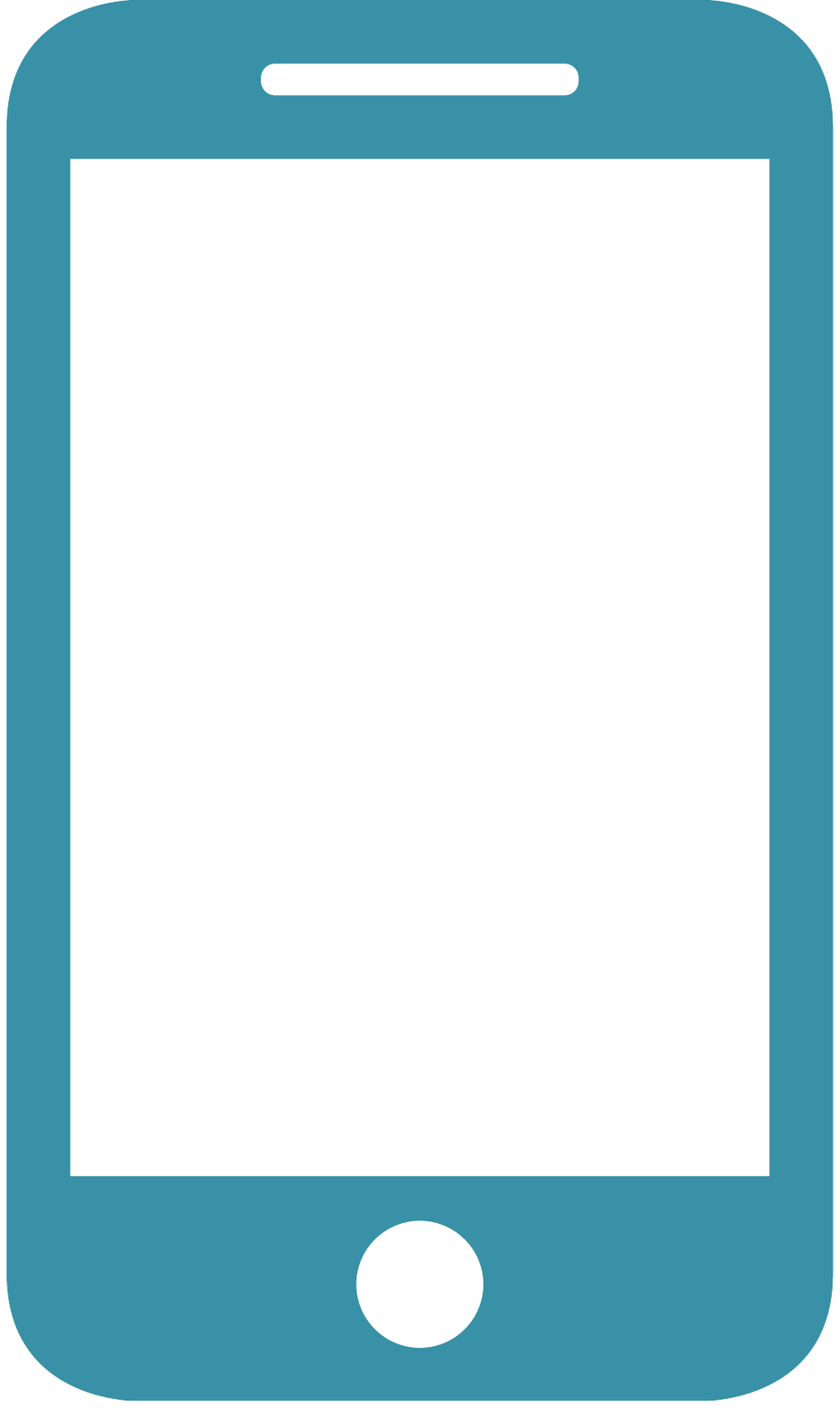}{figures/UE_col1}
\pgfdeclareimage[height=6mm]{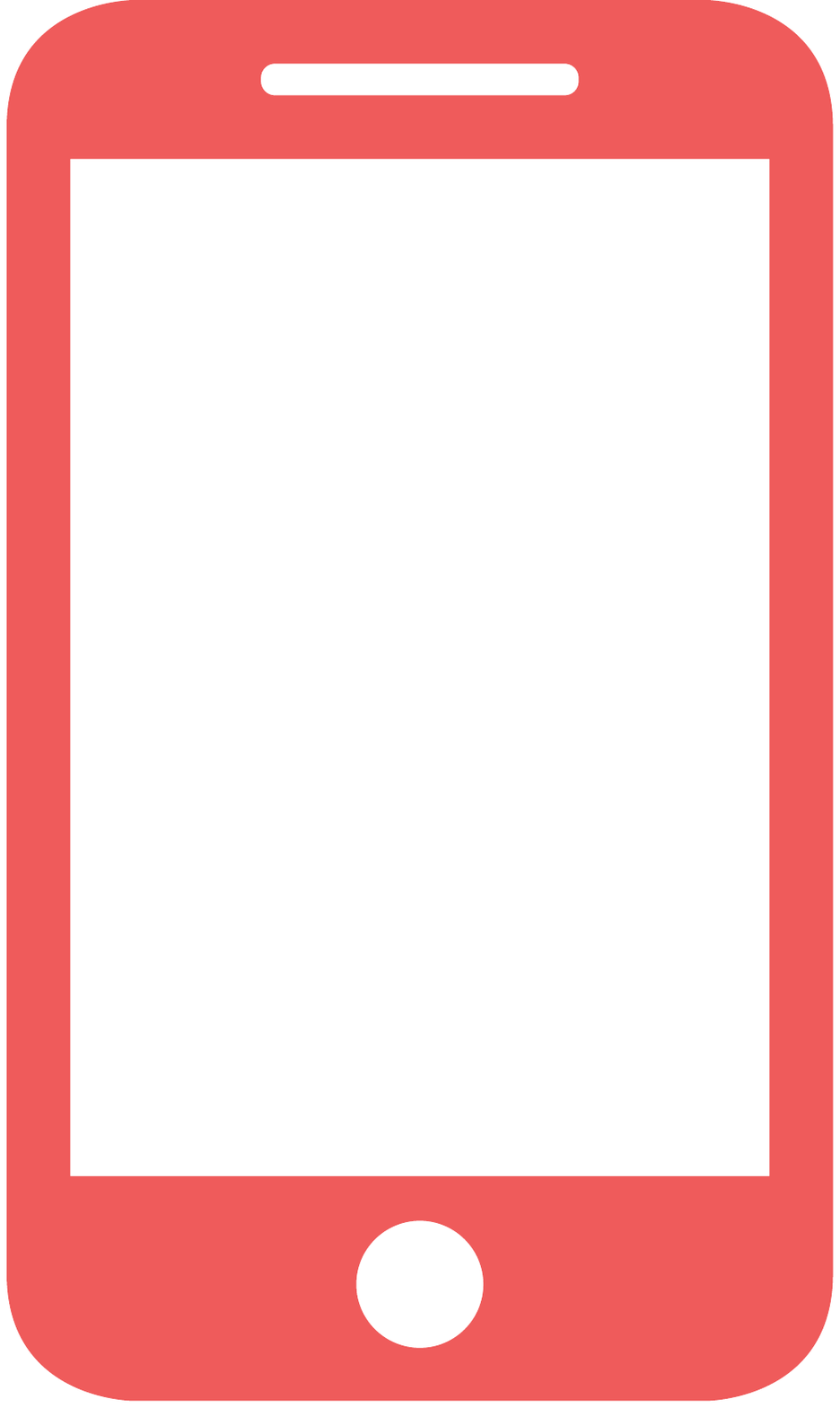}{figures/UE_col2}
\pgfdeclareimage[height=6mm]{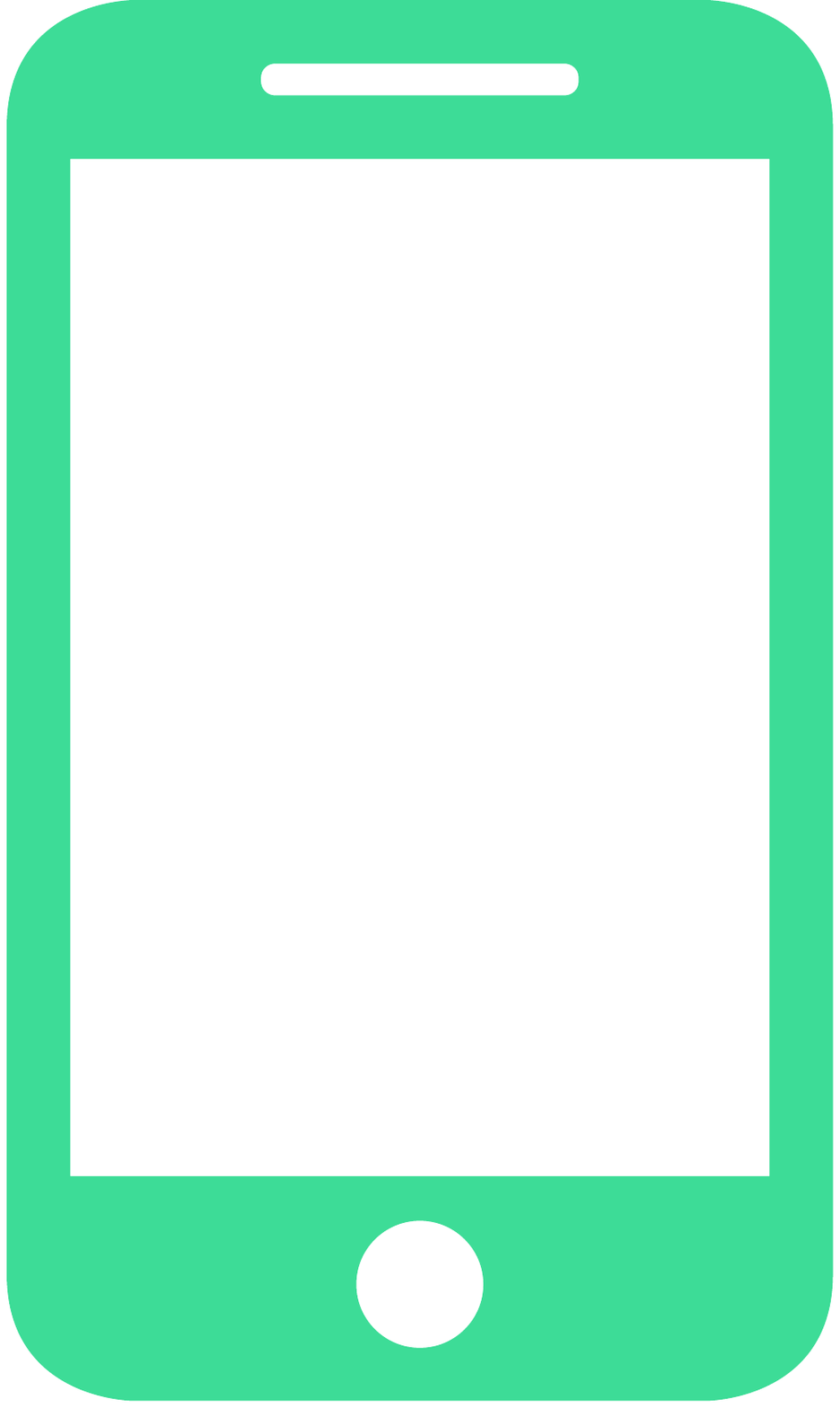}{figures/UE_col3}

\usetikzlibrary{calc}

\newdimen\d
\d=8mm;

\scriptsize


\draw node[align=center, anchor=north] (0_left) {};

\node[left of=0_left, node distance=2\d, align=center, anchor=center] (BS_1) {\pgfbox[center,bottom]{\pgfuseimage{BS}}};
\node[below of=0_left, node distance=2\d, align=center, anchor=center] (BS_2) {\pgfbox[center,bottom]{\pgfuseimage{BS}}};
\node[right of=0_left, node distance=2\d, align=center, anchor=center] (BS_3) {\pgfbox[center,bottom]{\pgfuseimage{BS}}};
\draw node[below of=BS_1, node distance=0.25\d, align=center, anchor=center] () {AP};
\node[left of=0_left, node distance=0.75\d, align=center, anchor=center] (UE_1) {\pgfbox[center,bottom]{\pgfuseimage{UE_col1}}};
\node[below of=UE_1, node distance=1\d, align=center, anchor=center] (UE_2) {\pgfbox[center,bottom]{\pgfuseimage{UE_col2}}};
\node[below of=0_left, node distance=0.5\d, align=center, anchor=center] (UE_3) {\pgfbox[center,bottom]{\pgfuseimage{UE_col1}}};
\node[right of=0_left, node distance=0.75\d, align=center, anchor=center] (UE_4) {\pgfbox[center,bottom]{\pgfuseimage{UE_col3}}};
\node[right of=UE_2,, node distance=1.5\d, align=center, anchor=center] (UE_5) {\pgfbox[center,bottom]{\pgfuseimage{UE_col2}}};

\draw node[above of=UE_1, node distance=0.45\d, align=center, anchor=center, xshift=-0.05\d] () {1};
\draw node[above of=UE_2, node distance=0.45\d, align=center, anchor=center, xshift=-0.05\d] () {2};
\draw node[above of=UE_3, node distance=0.45\d, align=center, anchor=center, xshift=-0.05\d] () {3};
\draw node[above of=UE_4, node distance=0.45\d, align=center, anchor=center, xshift=-0.05\d] () {4};
\draw node[above of=UE_5, node distance=0.45\d, align=center, anchor=center, xshift=-0.05\d] () {5};
\draw node[below of=UE_5, node distance=0.25\d, align=center, anchor=center] () {UE};

\draw node[right of=0_left, node distance=4\d, align=center, anchor=center] (0_right) {};

\draw node[right of=0_right, node distance=0\d, align=center, anchor=center] (DL) {};
\draw node[below of=DL, node distance=1\d, align=center, anchor=center] (UL) {};

\node[right of=DL, node distance=0.75\d, align=center, anchor=center] (UE_1_DL) {\pgfbox[center,bottom]{\pgfuseimage{UE_col1}}};
\node[right of=DL, node distance=2*0.75\d, align=center, anchor=center] (UE_2_DL) {\pgfbox[center,bottom]{\pgfuseimage{UE_col2}}};
\node[right of=DL, node distance=3*0.75\d, align=center, anchor=center] (UE_3_DL) {\pgfbox[center,bottom]{\pgfuseimage{UE_col1}}};
\node[right of=DL, node distance=5*0.75\d, align=center, anchor=center] (UE_5_DL) {\pgfbox[center,bottom]{\pgfuseimage{UE_col2}}};
\node[right of=UL, node distance=0.75\d, align=center, anchor=center] (UE_1_UL) {\pgfbox[center,bottom]{\pgfuseimage{UE_col1}}};
\node[right of=UL, node distance=3*0.75\d, align=center, anchor=center] (UE_3_UL) {\pgfbox[center,bottom]{\pgfuseimage{UE_col1}}};
\node[right of=UL, node distance=4*0.75\d, align=center, anchor=center] (UE_4_UL) {\pgfbox[center,bottom]{\pgfuseimage{UE_col3}}};

\draw node[above of=UE_1_DL, node distance=0.45\d, align=center, anchor=center, xshift=-0.05\d] () {1};
\draw node[above of=UE_2_DL, node distance=0.45\d, align=center, anchor=center, xshift=-0.05\d] () {2};
\draw node[above of=UE_3_DL, node distance=0.45\d, align=center, anchor=center, xshift=-0.05\d] () {3};
\draw node[above of=UE_5_DL, node distance=0.45\d, align=center, anchor=center, xshift=-0.05\d] () {5};
\draw node[above of=UE_1_UL, node distance=0.45\d, align=center, anchor=center, xshift=-0.05\d] () {1};
\draw node[above of=UE_3_UL, node distance=0.45\d, align=center, anchor=center, xshift=-0.05\d] () {3};
\draw node[above of=UE_4_UL, node distance=0.45\d, align=center, anchor=center, xshift=-0.05\d] () {4};
\draw node[above of=DL, node distance=0.45\d, align=center, anchor=center] () {DL:};
\draw node[above of=UL, node distance=0.45\d, align=center, anchor=center] () {UL:};

\end{tikzpicture}
\end{center}
\caption{System model with $\setK \! = \! \{ 1, 2, 3, 4, 5 \}$, $\setK^{\dl} \! = \! \{ 1, 2, 3, 5 \}$, $\setK^{\ul} \! = \! \{ 1, 3, 4 \}$, $\setK^{\dl\textnormal{-}\ul} \! = \! \{ 1, 3 \}$, $\setK^{\dl\textnormal{-only}} \! = \! \{ 2,5 \}$, and $\setK^{\ul\textnormal{-only}} \! = \! \{ 4 \}$.}
\label{fig:sys_mod}
\vspace{-2mm}
\end{figure}

We consider a cell-free massive MIMO system operating in TDD mode with channel reciprocity, where a set of APs $\setB$, each equipped with $M$ antennas, serves a set of UEs $\setK$, each equipped with $N$ antennas. Let $\H_{b,k} \in \Compl^{M \times N}$ denote the UL channel matrix between UE~$k \in \setK$ and AP~$b \in \setB$. We assume that both DL and UL data transmissions take place within a resource block. In this context, let $\setK^{\dl}$ and $\setK^{\ul}$ be the sets of DL and UL UEs, respectively, with $\setK^{\dl} \cup \setK^{\ul} = \setK$. The UEs served in both the DL and UL are referred to as \textit{DL-UL} UEs and are included in the set $\setK^{\dl\textnormal{-}\ul} \triangleq \setK^{\dl} \cap \setK^{\ul}$. Likewise, the UEs served only in the DL (resp. UL) are referred to as \textit{DL-only} (resp. \textit{UL-only}) UEs and are included in the set $\setK^{\dl\textnormal{-only}} \triangleq \setK \setminus \setK^{\ul}$ (resp. $\setK^{\ul\textnormal{-only}} \triangleq \setK \setminus \setK^{\dl}$). Without loss of generality, we assume a single DL and/or UL data stream per UE. An example of the system model is illustrated in Fig.~\ref{fig:sys_mod}.

\smallskip

\textit{\textbf{DL data transmission.}} Let $\w^{\dl}_{b,k} \in \Compl^{M \times 1}$ be the DL precoder used by AP~$b$ for UE~$k \in \setK^{\dl}$. The received signal at UE~$k$ is
\begin{align} \label{eq:y_k} 
\y^{\dl}_{k}  \triangleq \sum_{b \in \setB} \sum_{\bar{k} \in \setK^{\dl}} \H_{b,k}^{\herm} \w^{\dl}_{b,\bar{k}} d^{\dl}_{\bar{k}} + \z^{\dl}_{k} \in \Compl^{N \times 1}, 
\end{align}
where $d^{\dl}_k \sim \setC \setN (0,1)$ is the DL data symbol for UE~$k$ and $\z^{\dl}_{k} \sim \setC \setN (\0,\sigma_{\ue}^{2} \I_N)$ is the additive white Gaussian noise (AWGN) at UE~$k$. Subsequently, UE~$k$ uses its DL combiner $\v^{\dl}_{k} \in \Compl^{N \times 1}$ to obtain a soft estimate of ${d}^{\dl}_k$, i.e., $\hat{d}^{\dl}_k = (\v^{\dl}_{k})^\herm \y^{\dl}_{k}$. The corresponding DL signal-to-interference-plus-noise ratio (SINR) is given by
\begin{align} \label{eq:SINR_k_DL}
\gamma^{\dl}_{k} \triangleq  \frac{  |(\v^{\dl}_{k})^{\herm} \f_{k, k}|^{2}}{  \sum_{\bar k \in \setK^{\dl} \setminus \{k\}} | (\v^{\dl}_{k})^{\herm} \f_{k,\bar k}|^{2} + \sigma_{\ue}^{2} \| \v^{\dl}_{k} \|^{2}},
\end{align}
with $\f_{k,\bar k} \triangleq \sum_{b \in \setB}\H_{b,k}^{\herm} \w^{\dl}_{b,\bar k}$.

\smallskip

\textit{\textbf{UL data transmission.}} Let $\v^{\ul}_{k} \in \Compl^{N \times 1}$ be the UL precoder used by UE~$k \in \setK^{\ul}$. The received signal at AP~$b$ is
\begin{align} \label{eq:y_kul}
\y^{\ul}_{b} \triangleq \sum_{\bar{k} \in \setK^{\ul}} \H_{b,\bar k} \v^{\ul}_{\bar{k}} d^{\ul}_{\bar{k}} + \z^{\ul}_{b} \in \Compl^{M \times 1},
\end{align}
where $d^{\ul}_k \sim \setC \setN (0,1)$ is the UL data symbol of UE~$k$ and $\z_{b} \sim \setC \setN (\0,\sigma_{\bs}^{2} \I_M)$ is the AWGN at AP~$b$. Subsequently, each AP~$b$ uses its UL combiner $\w_{b,k}^{\ul} \in \Compl^{M \times 1}$ for UE~$k$ and the APs collectively obtain a soft estimate of ${d}^{\ul}_k$, i.e., $\hat{d}^{\ul}_k = \sum_{b \in \setB} (\w^{\ul}_{b,k})^\herm \y^{\ul}_{b}$. The corresponding UL SINR is given~by 
\begin{align} \label{eq:SINR_k_UL}
\gamma^{\ul}_{k} \triangleq \frac{|\g_{k, k}^{\herm} \v^{\ul}_{k} |^{2}}{\sum_{\bar k \in \setK^{\ul}\setminus \{k\} } | \g_{\bar k, k}^{\herm}  \v^{\ul}_{\bar k}|^{2} +  \sum_{b \in \setB} \sigma_{\bs}^{2} \| \w^{\ul}_{b,k} \|^{2}}.
\end{align}
with $\g_{\bar k, k} \triangleq \sum_{b \in \setB}\H_{b,\bar k}^{\herm} \w^{\ul}_{b, k}$.

\smallskip

Finally, the DL and UL sum rates (measured in bps/Hz) are given by $R^{\mathrm x} \triangleq \sum_{k \in \setK^{\mathrm x}} \log_{2}(1 + \gamma^{\mathrm x}_{k})$, with $\mathrm{x} \in \{\textnormal{DL}, \textnormal{UL} \}$. These rates represent upper bounds on the system performance (see \cite{Atz21,Gou24}) and are used as main performance metrics in Section~\ref{sec:NUM}. Next, we present the proposed combined DL-UL distributed beamforming designs with perfect CSI and IBT in Sections~\ref{sec:MSEJDU} and~\ref{sec:Dis_Imp_BiT}, respectively.

\section{Beamforming Design with Perfect CSI} \label{sec:MSEJDU}

The proposed combined DL-UL distributed beamforming designs are based on the following principles:

\textit{1)}~For each DL-UL UE, each AP adopts the same beamformer (up to a scaling) as precoder for the DL data transmission and as combiner for the UL data transmission. Similarly, each DL-UL UE adopts the same beamformer (up to a scaling) as combiner for the DL data transmission and as precoder for the UL data transmission. This approach exploits the structural similarity between the DL and UL beamformers resulting from separate DL and UL designs~\cite{Gou21a}.

\textit{2)}~Adopting the same beamformers (up to a scaling) for the DL and UL data transmissions enables using a single IBT procedure for all the UEs. This leads to a reduced IBT overhead compared with separate DL and UL designs at the cost of extra interference due to the DL-only or UL-only UEs.

To obtain a fully distributed beamforming design at the APs, we consider the minimization of the sum MSE, which provides some in-built fairness among the UEs (and can be further tuned by means of UE-specific weights)~\cite{Gou24}. In the rest of this section, we present the proposed DL-UL distributed beamforming design assuming perfect global CSI. The case with imperfect CSI obtained via IBT is described in Section~\ref{sec:Dis_Imp_BiT}.

\subsection{UE-Specific Beamforming Design}\label{sec:dis_bf_design}

Assuming that all the UEs are served solely in the DL, the precoders $\{\w^{\dl}_{b, k} \}$ at the APs and the combiners $\{\v^{\dl}_{k}\}$ at the UEs are obtained by solving
\begin{align} \label{eq:nMSEParDL}
\begin{array}{cl}
\underset{{\{\w^{\dl}_{b, k}, \v^{\dl}_{k}\}}}{\mathrm{minimize}} & \displaystyle \sum_{k \in \setK}  \mse^{\dl}_{k} \\
\mathrm{s.t.} & \displaystyle \sum_{k \in \setK} \| \w^{\dl}_{b,k}\|^{2} \leq \rho_{\bs}, 
\end{array}
\end{align}
where $\rho_{\bs}$ is the maximum transmit power of each AP and $\mse^{\dl}_{k} \triangleq \Exp \big[ |\hat d^{\dl}_{k} - d^{\dl}_{k}|^{2} \big]$. Since the objective is convex in $\{\w^{\dl}_{b, k}\}$ for fixed $\{\v^{\dl}_{k}\}$ and vice versa, we use alternating optimization to solve \eqref{eq:nMSEParDL} (as in~\cite{Atz21}).

\smallskip

\textit{\textbf{Optimization of $\v^{\dl}_{k}$.}} For fixed $\{\w^{\dl}_{b, k}\}$, the optimal $\v^{\dl}_{k}$ is computed by setting $\nabla_{\v_{k}^{\dl}}\big(\sum_{k \in \setK} \mse_{k}^{\dl} \big)  = 0$, which yields
\begin{align} \label{eq:v_k}
\v^{\dl}_{k} = \bigg( \sum_{\bar k \in \setK^{\dl\textnormal{-}\ul} \cup {\setK^{\dl\textnormal{-only}}}  \cup \setK^{\ul\textnormal{-only}}}   \f_{k, \bar k}  \f_{k, \bar k}^{\herm} + \sigma_{\ue}^{2} \I_{N} \bigg)^{ -1} \f_{k, k}.
\end{align}

\smallskip

\textit{\textbf{Optimization of $\w^{\dl}_{b, k}$.}} For fixed $\{\v^{\dl}_{k}\}$, the optimal $\w^{\dl}_{b, k}$ is computed by setting $\nabla_{\w_{b,k}^{\dl}}\big(\sum_{k \in \setK} \mse_{k}^{\dl} \big)  = 0$, which yields
\begin{align} \label{eq:w_bk}
\w^{\dl}_{b,k} = (\Phib_{b b} + \lambda_{b} \I_{M})^{-1} (\h_{b,k}- \xib_{b,k}),
\end{align}
where $\Phib_{b \bar{b}} \triangleq \sum_{\bar k \in \setK^{\dl\textnormal{-}\ul} \cup \setK^{\dl\textnormal{-only}} \cup \setK^{\ul\textnormal{-only}}} \h_{b, \bar k} \h_{\bar b,  \bar k}^{\herm}$, 
$\lambda_{b}$ is the dual variable corresponding to AP~$b$'s transmit power constraint, $\h_{b,k} \triangleq \H_{b,k} \v^{\dl}_{k}$, and $\xib_{b,k} \triangleq \sum_{\bar{b} \in \setB \setminus \{b\}} \Phib_{b \bar{b}} \w^{\dl}_{\bar{b},k}$. Furthermore, to ensure the global convergence of \eqref{eq:nMSEParDL} with respect to $\w^{\dl}_{b, k}$, each AP updates $\w^{\dl}_{b, k}$ in \eqref{eq:w_bk} via a best-response (BR) method, while the term $\xib_{b,k}$ needs to be acquired from the other APs as detailed in~\cite{Atz21}.

Incorporating all the UEs into the DL beamforming design facilitates the reuse of the DL beamformers for the UL data transmission. Specifically, for a given $k \in \setK^{\ul}$, \eqref{eq:v_k} can be reused as UL precoder at the UE and \eqref{eq:w_bk} can be reused as UL combiner at AP~$b$. However, by this approach, the APs unnecessarily aim to cancel interference due to the UL-only UEs in the DL and the DL-only UEs in the UL (reflected by the summations over $\setK^{\ul\textnormal{-only}}$ and $\setK^{\dl\textnormal{-only}}$, respectively, in $\Phib_{b \bar{b}}$). Likewise, the UEs unnecessarily aim to cancel the effective interference due to the UL-only UE beamformers in the DL and the DL-only UE beamformers in the UL (reflected by the summations over $\setK^{\ul\textnormal{-only}}$ and $\setK^{\dl\textnormal{-only}}$, respectively, in \eqref{eq:v_k}).

The DL beamformers in \eqref{eq:w_bk} and \eqref{eq:v_k} can be respectively enhanced for the DL data transmission and reused for the UL data transmissions to satisfy the transmit power constraints at the APs and UEs. For the DL data transmission, \eqref{eq:w_bk} corresponding to UE~$k \in \mathcal{K}^{\text{DL}}$ is scaled to meet the AP transmit power constraint after discarding the UL-only UEs, i.e., $a_b \mathbf{w}^{\text{DL}}_{b, k}$, with $a_b \triangleq {\rho_{\text{BS}}}/\sum_{k \in \mathcal{K}^{\text{DL}}} \|\mathbf{w}^{\text{DL}}_{b, k}\|^2$. For the UL data transmission, \eqref{eq:v_k} corresponding to UE~$k \in \mathcal{K}^{\text{UL}}$ is reused as UL precoder and scaled to meet the UE transmit power constraint, i.e., $\mathbf{v}^{\text{UL}}_{k} = \sqrt{\rho_{\text{UE}}}{\mathbf{v}^{\text{DL}}_{k}}/{\| \mathbf{v}^{\text{DL}}_{k} \|}$, where $\rho_{\text{UE}}$ is the maximum transmit power of each UE.

\subsection{Beamforming Design with UE Pairing}\label{sec:dis_ue_grp}

To further reduce the number of precoders at each AP (and thus the IBT overhead), DL-only UEs can be paired with UL-only UEs and each pair is assigned a common DL multicast precoder during the IBT. Let $\setG \triangleq \big\{ 1, \ldots, \max ( |\setK^{\dl}|, |\setK^{\ul}| ) \big\}$ and let $P_g \triangleq (a_{g}, b_{g})$ denote an ordered pair of DL and UL UEs, with $g \in \setG$, $a_{g} \in \setK^{\dl}$, and $b_{g} \in \setK^{\ul}$. In this setting, we have $a_{g} = b_{g} = k$ if $k \in \setK^{\dl\textnormal{-}\ul}$ or $a_{g} = k \neq \bar{k} = b_{g}$ if $k \in \setK^{\dl\textnormal{-only}}$ and $\bar{k} \in \setK^{\ul\textnormal{-only}}$. Moreover, when $|\setK^{\dl}| \neq |\setK^{\ul}|$, the remaining DL-only or UL-only UEs are paired with a phantom UE indicated by the index $0$. In the example of system model in Fig.~\ref{fig:sys_mod}, we have $P_{1} = (1,1)$, $P_{2} = (2,4)$, $P_{3} = (3,3)$, and $P_{4} = (5,0)$. By this approach, the precoders  $\{\w_{b,g}^{\dl}\}$ at the APs and the combiners $\{ \v_k \}$ at the UEs are obtained by solving the following virtual multicast problem:
\begin{align} \label{eq:gMSEParDL}
\begin{array}{cl}
\underset{{\{\w^{\dl}_{b, g}, \v^{\dl}_{k} \}}}{\mathrm{minimize}} & \displaystyle  \sum_{k \in \setK} \gmse^{\dl}_{k} \\
\mathrm{s.t.} & \displaystyle  \sum_{g \in \setG} \| \w^{\dl}_{b,g}\|^{2} \leq \rho_{\bs},  
\end{array}
\end{align}
with $\gmse^{\dl}_{k} \triangleq \Exp \big[ |(\v^{\dl}_{k})^{\herm} \y^{\dlgrp}_{k} - d^{\dl}_{g_k}|^{2} \big]$, where $g_k$ represents the index of the pair containing UE~$k$ and $\y^{\dlgrp}_{k} \triangleq \sum_{b \in \setB} \sum_{\bar{g} \in \setG} \H_{b,k}^{\herm} \w^{\dl}_{b,\bar g} d^{\dl}_{\bar g} + \z^{\dl}_{k}$. Similar to~\eqref{eq:nMSEParDL}, the objective in~\eqref{eq:gMSEParDL}  is convex in $\{\w^{\dl}_{b, g}\}$ for fixed $\{\v^{\dl}_{k}\}$ and vice versa, so we use alternating optimization to solve \eqref{eq:gMSEParDL} (as in~\cite{Gou24}). 

\smallskip

\textit{\textbf{Optimization of $\v^{\dl}_{k}$.}} For fixed $\{\w^{\dl}_{b,g}\}$, the optimal $\v^{\dl}_{k}$ is computed as $\nabla_{\v_{k}^{\dl}}\big(\sum_{k \in \setK}       \gmse_{k}^{\dl}   \big) = 0$, which yields
\begin{align} \label{eq:v_k_g}
\v^{\dl}_{k} \! = \! \bigg( \sum_{\bar g \in \setG} (\mathbbm{1}_{\bar g} \m_{k, \bar g} \m_{k,  \bar g}^{\herm} + \underbrace{ \bar {\mathbbm{1}}_{\bar g} \m_{k, \bar g} \m_{k, \bar g}^{\herm}}_{\textnormal{extra interference}}) \! + \sigma_{\ue}^{2} \I_{N}  \bigg)^{\!-1} \! \m_{k,g},
\end{align}
with $\m_{k,g} \triangleq \sum_{b \in \setB} \H_{b,k}^{\herm} \w^{\dl}_{b,g}$, $\bar {\mathbbm{1}}_{g} \triangleq 1 - {\mathbbm{1}}_{ g}$, and $\mathbbm{1}_{g} = 0$ if $\{P_g\} \cap \setK^{\dl} = \emptyset$ or $\{P_g\} \cap \setK^{\ul} = \emptyset$ (otherwise $\mathbbm{1}_{g} = 1$). This implies that extra interference exists in the DL (resp. UL) when a UL-only (resp. DL-only) UE is paired with a phantom UE.

\smallskip

\textbf{\textit{Optimization of $\w^{\dl}_{b, g}$.}} To reduce the IBT overhead, we use the gradient method to obtain the optimal $\w^{\dl}_{b, g}$ for a fixed $\{\v^{\dl}_{k}\}$ (as in~\cite{Gou24}). Accordingly, let us define \vspace{-0.5mm}
\begin{align} \label{eq:BSUd}
\delta \mathbf{\epsilon}_{b,g} & \triangleq \nabla_{\w_{b,g}^{\dl}}\bigg(\sum_{k \in \setK} \gmse_{k}^{\dl} \bigg) \\
& \label{eq:BSUd2} = - 2  \bigg( \sum_{k \in \setK_g} \h_{b, k} - \mathbf{\varrho}_{b,g} - \sum_{k \in \setK} \h_{b,k} \h_{b, k}^{\herm} \w_{b,g}^{\dl} \bigg),
\end{align}
where $\mathbf{\varrho}_{b,g} \triangleq \sum_{\bar b \in \setB \setminus \{b\}}\sum_{k \in \setK} \h_{b,k} \h_{\bar b, k}^{\herm} \w_{\bar b,g}^{\dl}$ needs to be acquired from the other APs. Then, the corresponding gradient \newpage \noindent update of $\w_{b,g}^{\dl}$ at iteration $i$ is
\begin{align}\label{eq:grd_bs_up}
\tilde{\w}_{b,g}^{\dl (i)}  \triangleq \w_{b,g}^{\dl (i-1)}  - \alpha_{\textnormal{\tiny{GB}}} \delta \mathbf{\epsilon}_{b,g},
\end{align}
where $\alpha_{\textnormal{\tiny{GB}}}$ is the step size. The precoders at AP~$b$~are then scaled to meet the per-AP transmit power constraints, i.e., ${\w}_{b,g}^{\dl(i)} = \bar a_b \tilde{\w}_{b,g}^{\dl(i)}$, such that   $ \bar a_b = \rho_{\bs}/  \|\sum_{g \in \setG} \tilde {\w}_{b,g}^{ \dl (i)} \|^2$. While this distributed beamforming design offers the advantage of reducing the IBT overhead and considering interference only from other UE pairs, it also incurs a beamforming loss due to unnecessary multicasting in the presence of UE pairs with DL-only and UL-only UEs. Finally,  $\v^{\dl}_{k}$ in \eqref{eq:v_k_g} and $\w^{\dl}_{b,g}$ in  \eqref{eq:grd_bs_up}  are scaled for the DL and UL data transmissions as discussed in Section~\ref{sec:dis_bf_design}.

\section{Beamforming Design with IBT}\label{sec:Dis_Imp_BiT}

The distributed beamforming design described in Section~\ref{sec:dis_bf_design} can be carried out at each AP and UE by means of IBT. At each IBT iteration, the DL precoders and combiners for all the UEs are updated at each AP and UE, respectively, via precoded pilots.

\subsection{UE-Specific Beamforming Design}\label{sec:Dis_imp}

Let $\p_k \in \Compl^{\tau\times 1}$ be the pilot assigned to UE~$k$, such that $\p_k^{\herm} \p_{\bar k}= \tau$ if $k = \bar k$ and  $\p_k^{\herm} \p_{\bar k}= 0$ otherwise.

\smallskip

\textit{\textbf{Computation of $\v^{\dl}_{k}$.}} To enable the computation of $\v^{\dl}_{k}$ in \eqref{eq:v_k} at each UE, each AP~$b$ transmits the precoded DL pilots $\X_{b}^{\dl} \triangleq \sum_{k \in \setK} \w^{\dl}_{b,k} \p_{k}^{\herm}$. The received signal at UE~$k$ from all the APs is
\begin{align}\label{eq:y_k_dl}
    \Y_{k}^{\dl} \triangleq  \sum_{b \in \setB} \sum_{\bar k \in \setK}    \H_{b, k}^{\herm} \w^{\dl}_{b,\bar k} \p_{\bar k}^{\herm}  +  \Z_{k}^{\dl} \in \Compl^{N \times \tau},
\end{align}
where $ \Z_{k}^{\dl}$ is the AWGN at UE~$k$. Then, UE~$k$ computes $\v^{\dl}_{k}$~as
\begin{align}\label{eq:v_k_dl_comp}
   \v^{\dl}_{k} \simeq \big( \Y_{k}^{\dl} (\Y_{k}^{\dl})^{\herm} \big)^{\dagger} \Y_{k}^{\dl} \p_{k},
\end{align}
where $(\cdot)^{\dagger}$ is the pseudoinverse operator. Note that the above $\v^{\dl}_{k}$ converges to~\eqref{eq:v_k} as $\tau \rightarrow \infty$.

\smallskip

\textit{\textbf{Computation of $\w^{\dl}_{b,k}$.}} To enable the computation of $\w^{\dl}_{b,k}$ in \eqref{eq:w_bk} at each AP, each UE~$k$ transmits the precoded UL pilot $\X_{k}^{\ulA} \triangleq \frac{1}{\sqrt{\beta}}\v^{\dl}_{k} \p_{k}^{\herm}$, where $\beta$ is a scaling factor that ensures the per-UE transmit power constraint (equal for all the UEs). The received signal at AP~$b$ from all the UEs is
\begin{align}\label{eq:yb_1}
    \Y_{b}^{\ulA} \triangleq \frac{1}{\sqrt{\beta}} \sum_{k \in \setK} \H_{b, k}\v^{\dl}_{k}\p_{k}^{\herm} +  \Z_{b}^{\ulA} \in \Compl^{M \times \tau},
\end{align}
where $\Z_{b}^{\ulA}$ is the AWGN at AP~$b$. Then, to reconstruct $\xib_{b,k}$ in \eqref{eq:w_bk}, each UE~$k$ transmits an extra OTA signal obtained by precoding $\Y_{k}^{\dl}$ with $\v^{\dl}_{k}(\v^{\dl}_{k})^{\herm}$, i.e., $\X_{k}^{\ulB} \triangleq  \frac{1}{\sqrt{\beta}} \v^{\dl}_{k}(\v^{\dl}_{k})^{\herm} \Y_{k}^{\dl}$. The corresponding received signal at AP~$b$~is
\begin{align}\label{eq:yb_2}
     \Y_{b}^{\ulB} \triangleq \frac{1}{\sqrt{\beta}} \sum_{k \in \setK} \H_{b, k}\v^{\dl}_{k}(\v^{\dl}_{k})^{\herm} \Y_{k}^{\dl}  +   \Z_{b}^{\ulB} \in \Compl^{M \times \tau},
\end{align}
where $\Z_{b}^{\ulB}$ is the AWGN at AP~$b$. From \eqref{eq:yb_1} and \eqref{eq:yb_2}, $\w^{\dl}_{b,k}$ in~\eqref{eq:w_bk} is approximated as 
\begin{align} \label{eq:w_bk_imp_ota}
\w_{b,k}^{\dl} & \simeq  \big( \Y_{b}^{\ulA} (\Y_{b}^{\ulA})^{\herm}  + \tau ( \beta \lambda_{b} - \sigma_{\bs}^{2}) \I_{M} \big)^{\dagger} \Big( \Y_{b}^{\ulA} \big( \sqrt{\beta} \p_{k} \nonumber \\
& \phantom{=} \ + (\Y_{b}^{\ulA})^{\herm}\w_{b,k}^{\dl} \big) - \sqrt{\beta}\Y_{b}^{\ulB} \p_{k} - \tau \sigma_{\bs}^{2} \w_{b,k}^{\dl} \Big),
\end{align}
which is updated via the BR method~\cite{Atz21}. As $\tau \rightarrow \infty$, $\w_{b,k}^{\dl}$ in \eqref{eq:w_bk_imp_ota} converges to \eqref{eq:w_bk}.

\subsection{Beamforming Design with UE Pairing}\label{sec:Dis_imp_UEGrp}

The distributed beamforming design with UE pairing detailed in Section~\ref{sec:dis_ue_grp} can be carried out in a similar way as in Section~\ref{sec:Dis_imp}. Let $\p_g \in \Compl^{\tau_{\tg} \times 1}$ be the pilot assigned to pair~$g \in \setG$, such that $\p_g^{\herm} \p_{\bar g}= \tau_{\tg}$ if $g = \bar g$ and $\p_g^{\herm} \p_{\bar g}= 0$ otherwise.

\smallskip

\textit{\textbf{Computation of $\v^{\dl}_{k}$.}} To enable the computation of $\v^{\dl}_{k}$ in \eqref{eq:v_k_g} at each UE, each AP~$b$ transmits the precoded DL pilots $\X_{b}^{\dl} \triangleq \sum_{g \in \setG} \w_{b,g} \p_{g}^{\herm}$. The received signal at UE~$k$ from all the APs is
\begin{align}\label{eq:y_k_dl_grp}
    \Y_{k}^{\dl} \triangleq \sum_{b \in \setB} \sum_{g \in \setG} \H_{b, k}^{\herm} \w_{b,g} \p_g^{\herm} + \Z_{k}^{\dl} \in \Compl^{N \times \tau_{\tg}}.
\end{align}
Based on \eqref{eq:y_k_dl_grp}, UE~$k$ in $P_g$ computes $\v^{\dl}_{k}$ as
\begin{align}\label{eq:v_k_dl_grp_comp}
    \v^{\dl}_{k} \simeq (\Y_{k}^{\dl} (\Y_{k}^{\dl})^{\herm})^{\dagger} \Y_{k}^{\dl} \p_{g_k}. 
\end{align}
Note that the above $\v^{\dl}_{k}$ converges to~\eqref{eq:v_k_g} as $\tau_{\tg} \rightarrow \infty$.
\smallskip

\smallskip

\textit{\textbf{Computation of $\w_{b,g}^{\dl}$.}} The computation of $\w^{\dl}_{b,k}$ in \eqref{eq:w_bk} with UE pairing requires the gradient in \eqref{eq:BSUd} at each AP. To allow AP~$b$ to reconstruct $\mathbf{\varrho}_{b,g}$ in \eqref{eq:BSUd2}, each UE~$k$ in $P_g$ transmits the precoded UL pilots $\X_{k}^{\ulA} = \frac{1}{\sqrt{\beta}}\v^{\dl}_{k} \p_{g_k}^{\herm}$ and the received signal at AP~$b$ is
\begin{align}\label{eq:ybg_1}
    \Y_{b}^{\ulA} = \frac{1}{\sqrt{\beta}} \sum_{k \in \setK} \H_{b, k}\v^{\dl}_{k} \p_{g_k}^{\herm} +  \Z_{b}^{\ulA} \in \Compl^{M \times \tau_{\tg}}.
\end{align}
Then, each UE~$k$ transmits an extra OTA signal as in Section~\ref{sec:Dis_imp} and the corresponding received signal at AP~$b$ is
\begin{align}\label{eq:ybg_2}
     \Y_{b}^{\ulB} = \frac{1}{\sqrt{\beta}} \sum_{k \in \setK} \H_{b, k}\v^{\dl}_{k}(\v^{\dl}_{k})^{\herm} \Y_{k}^{\dl} + \Z_{b}^{\ulB} \in \Compl^{M \times \tau_{\tg}}.
\end{align}
Finally, from \eqref{eq:ybg_1} and \eqref{eq:ybg_2}, AP~$b$ obtains
\begin{align}\label{eq:gd_comp}
\delta \mathbf{\epsilon}_{b,g} \simeq \frac{2}{\tau_{\tg}\sqrt{\beta}} (\Y_{b}^{\ulA} - \Y_{b}^{\ulB} ) \p_g,
\end{align}
after which it computes $\tilde \w_{b,g}^{\dl}$ as in \eqref{eq:grd_bs_up} and obtains $\w_{b,g}^{\dl}$ as in Section~\ref{sec:dis_ue_grp}. Note that the above $\delta \mathbf{\epsilon}_{b,g}$ converges to \eqref{eq:BSUd2} as $\tau_{\tg} \rightarrow \infty$. The distributed beamforming design with UE pairing can be implemented using pair-specific pilots, which allows to reduce the IBT overhead compared to the UE-specific design, especially when the number of DL-only and UL-only UEs is large. However, the UE pairing may introduce extra interference, as both paired UEs share a compromised beamforming strategy. 

\subsection{Implementation Details}

\begin{figure}
\begin{center}
\begin{tikzpicture}
\usetikzlibrary{calc}
\newdimen\d
\d=7mm;
\tiny
\tikzstyle{block} = [rectangle, draw, minimum width=\d, text centered, minimum height=1.2\d]
\node [block, minimum width=5mm, node distance=6mm, fill=yellow] (slot0) {{UL-1}};
\node [block, right of=slot0, minimum width=5mm, node distance=5mm, fill=green] (slot1) {DL};
\node [block, right of=slot1, minimum width=5mm, node distance=5mm, fill=yellow] (slot2) {UL-1};
\node [block, right of=slot2, minimum width=5mm, node distance=5mm, fill=pink] (slot3) {UL-2};
\node [block, right of=slot3, minimum width=10mm, node distance=7.5mm, fill=gray] (slot4) {UL-data};
\node [block, right of=slot4, minimum width=10mm, node distance=10mm, fill=gray] (slot5) {DL-data};
\node [block, right of=slot5, minimum width=5mm, node distance=7.5mm, fill=green] (slot6) {DL};
\node [block, right of=slot6, minimum width=5mm, node distance=5mm, fill=yellow] (slot7) {UL-1};
\node [block, right of=slot7, minimum width=5mm, node distance=5mm, fill=pink] (slot8) {UL-2};
\node [block, right of=slot8, minimum width=10mm, node distance=7.5mm, fill=gray] (slot9) {UL-data};
\node [block, right of=slot9, minimum width=10mm, node distance=10mm, fill=gray] (slot10) {DL-data};
\node [block, right of=slot10, minimum width=10mm, node distance=10mm,] (slot11) {$\cdots$};

\tiny
\draw [decorate, decoration={brace, raise=8.9mm, amplitude=1.5mm}] ($(slot0.south west)+(20mm,0mm)$) -- ($(slot5.south east)+(-0.3mm,0mm)$) node [xshift=-10mm, yshift=12mm, align=center, text width=2.5cm] {UL and DL data};
\draw [decorate, decoration={brace, raise=8.9mm, amplitude=1.5mm}] ($(slot0.south west)+(40mm,0mm)$) -- ($(slot5.south east)+(15mm,0mm)$) node [xshift=-8mm, yshift=12mm, align=center, text width=2.5cm] {Single IBT iteration};
\draw [decorate, decoration={brace, mirror, raise=0.5mm, amplitude=1.5mm}] ($(slot0.south west)+(0mm,0mm)$) -- ($(slot5.south east)+(-0.3mm,0mm)$) node [xshift=-19.5mm, yshift=-3.5mm, align=center, text width=2.5cm] {Resource block-1};
\draw [decorate, decoration={brace, mirror, raise=0.5mm, amplitude=1.5mm}] ($(slot6.south west)+(0.3mm,0mm)$) -- ($(slot10.south east)+(0mm,0mm)$) node [xshift=-17mm, yshift=-3.5mm, align=center, text width=2.5cm] {Resource block-2};
\end{tikzpicture}
\end{center}
\vspace{-2mm}
\caption{Resource blocks containing UL and DL resources for both data and IBT signalling.}
\label{fig:BiTframe}
\end{figure}
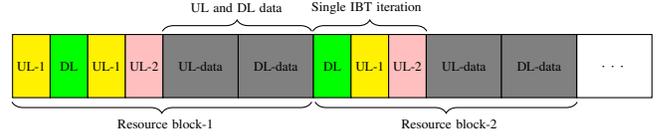

We adopt a strategy involving a single update of the combiners at all the UEs using a DL signal, and a single update of the precoders at each AP using the UL-1 and UL-2 signals per resource block (the DL signals and the UL-1 and UL-2 signals can be co-located with the DL and UL data, respectively). This approach allows for the proposed distributed beamforming design to be an integral part of the IBT process, as depicted in Fig.~\ref{fig:BiTframe}. However, in the first resource block, there is an extra UL-1 signaling, where the UEs initialize the combiners for the precoded pilot transmission. Afterward, each resource block contains the DL, UL-1, and UL-2 signals for IBT. The implementation of the proposed combined DL-UL distributed beamforming design is summarized in Algorithm~\ref{alg:disGd}.

\section{Numerical Results and Discussion} \label{sec:NUM}

We consider $B=25$ APs, each equipped with $M=8$ antennas, serving $|\setK|=32$ UEs (unless otherwise stated), each equipped with $N=4$ antennas. The APs are placed on a square grid with an area of $100 \times 100$~m$^2$ and distance between adjacent APs of $20$~m. We assume a time-varying channel model where the channel at resource block $t+1$ is related to the channel at resource block $t$ as (see Fig.~\ref{fig:BiTframe})~\cite{Mon06}
\begin{align}\label{eq:chMod}
    \H_{b,k}[t+1] = \kappa \H_{b,k}[t] + \sqrt{1-\kappa^2} \tilde{\H}_{b, k},
\end{align}
with $\kappa \in (0,1]$ and $\mathrm{vec}(\tilde{\H}_{b,k}) \sim \setC \setN (\0, \delta_{b,k} \I_{MN})$. Here, $\delta_{b,k} \triangleq -30.5-36.7\log_{10} (d_{b,k})$~[dB] is the large-scale fading coefficient, where $d_{b,k}$~[m] denotes the distance between AP~$b$ and UE~$k$. We consider a carrier frequency of $2.5$~GHz, a resource block duration of $5$~ms, and $\kappa=0.967$, where the latter corresponds to a UE mobility of $5$~km/h (walking speed). The maximum transmit power for both pilots and data is $\rho_{\bs} = 30$~dBm at the APs and $\rho_{\ue} = 20$~dBm at the UEs, whereas the AWGN power at both the APs and UEs is fixed to $\sigma_{\bs}^{2} = \sigma_{\ue}^{2} = -95$~dBm. Finally, the average effective DL-UL sum rate (in bps/Hz) at resource block~$t$ is computed as
$R(t) = (1- r_{\textrm{IBT}}/r_{\textrm{tot}}) \big(R^{\dl}(t) + R^{\ul}(t) \big)/2$.
Here $r_{\textrm{IBT}}$ and $r_{\textrm{tot}}$ are IBT resources and the total number of resources for data and IBT in a resource block, respectively. The minimum number of orthogonal resources for a single IBT iteration for the proposed and reference methods are given in Table~\ref{tab:resources}.

\begin{figure}[t]
\vspace{-4mm}
\begin{algorithm}[H]
\footnotesize
\begin{spacing}{1.15}
\textbf{Data:} Pilots $\{\p_{k}\}_{k \in \setK}$  or $\{\p_{g}\}_{g \in \setG}$.\\
\textbf{Initialization:} Combiners $\{\v^{\dl}_{k}\}_{k \in \setK}$.\\
\textbf{UL-1:} Each UE~$k$ transmits $\X_{k}^{\ulA}$ with $\p_{k}$ or $\p_{g_k}$, and each AP~$b$ receives $\Y_{b}^{\ulA}$ in \eqref{eq:yb_1} or \eqref{eq:ybg_1}, respectively. Then, compute $\w^{\dl}_{b,k}$ in \eqref{eq:w_bk_imp_ota} and update it via BR or $\delta \mathbf{\epsilon}_{b,g}$ in \eqref{eq:gd_comp} to obtain $\w_{b,g}^{\dl}$ as in Section~\ref{sec:dis_ue_grp}, with $\Y_{b}^{\ulB} = \0$. \\
\textbf{For each resource block}, \textbf{do:}
\begin{itemize}[]
\item[1)] \textbf{DL:} Each AP~$b$ transmits $\X_{b}^{\dl}$ with $\p_{k}$ or $\p_{g}$, and each UE~$k$ receives $\Y_{k}^{\dl}$ in \eqref{eq:y_k_dl} or \eqref{eq:y_k_dl_grp}, respectively, and computes $\v^{\dl}_{k}$ in \eqref{eq:v_k_dl_comp}~or~\eqref{eq:v_k_dl_grp_comp}.
\item[2)] \textbf{UL-1:}  Each UE~$k$ transmits $\X_{k}^{\ulA}$ with $\p_{k}$ or $\p_{g_k}$ and each AP~$b$ receives $\Y_{b}^{\ulA}$ in \eqref{eq:yb_1} or \eqref{eq:ybg_1}, respectively.
\item[3)] \textbf{UL-2:}  Each UE~$k$ transmits $\X_{k}^{\ulB}$ with \eqref{eq:y_k_dl} or \eqref{eq:y_k_dl_grp}, and each AP~$b$ receives $\Y_{b}^{\ulB}$ in \eqref{eq:yb_2} or \eqref{eq:ybg_2}, respectively. Then, compute $\w^{\dl}_{b,k}$ in \eqref{eq:w_bk_imp_ota} and update it via BR or $\delta \mathbf{\epsilon}_{b,g}$ in \eqref{eq:gd_comp} to obtain $\w_{b,g}^{\dl}$ as in Section~\ref{sec:dis_ue_grp}.
\item[4)] \textbf{UL data:} The DL beamformers are reused for the UL data transmission after scaling (see Section~\ref{sec:dis_bf_design}).
\item[5)] \textbf{DL data:}  The DL precoders are scaled for the DL data transmission (see Section~\ref{sec:dis_bf_design}).
\end{itemize}
\end{spacing}
\caption{(Combined DL-UL distr. beamforming design)} \label{alg:disGd}
\end{algorithm}
\vspace{-9mm}
\end{figure}

\addtolength{\tabcolsep}{-0.2em}
\begin{table*}[t]
\centering
\footnotesize
\begin{tabular}{|c|c|c|c|c|c|c|c|}
    \hline
    \textbf{Algorithm} & {Centralized} & {Sep. OTA}  & {Sep. local}  & {Comb. OTA}  & {Comb. local}  & {Comb. paired} OTA & {Comb. paired} local\\
    \hline
    $r_{\textrm{IBT}}$ & $ |\setK| N + |\setK^{\dl}| + |\setK^{\ul}| $ & $3(|\setK^{\dl}| + |\setK^{\ul}|)$ & $2(|\setK^{\dl}| + |\setK^{\ul}|)$ & $3 |\setK|$ & $2|\setK|$ & $3 \max(|\setK^{\dl}|, |\setK^{\ul}|)$ & $2 \max(|\setK^{\dl}|, |\setK^{\ul}|)$\\
    \hline
\end{tabular}
\caption{Minimum orthogonal resources required for a single IBT iteration in each resource block for the proposed and reference methods.}
\label{tab:resources}
\vspace{-2mm}
\end{table*}

\begin{figure}[t!]
\begin{center}
\input{figures/fig1.tex}
\vspace{-2mm}
\caption{Effective DL-UL sum rate vs. resource block number.}
\label{fig:rateVst}
\end{center}
\vspace{-4mm}
\end{figure}
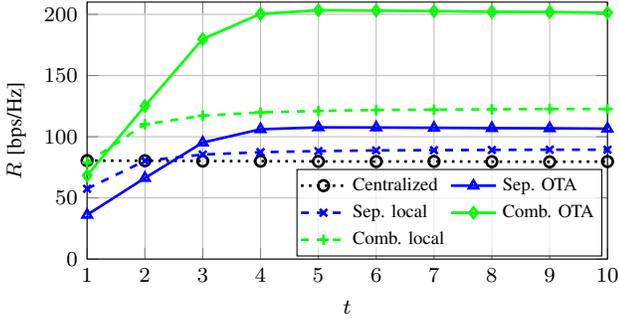

Fig.~\ref{fig:rateVst} plots the effective DL-UL sum rate against the resource block number, considering a time-varying channel as in \eqref{eq:chMod} with random initialization at $t=0$, for $r_{\textrm{tot}}=300$. The proposed combined DL-UL distributed beamforming design, referred to as \textit{comb. OTA}, outperforms all the other methods ($\setK^{\dl}=\setK^{\ul}$), with the combiner and precoder computed as in \eqref{eq:v_k_dl_comp} and \eqref{eq:w_bk_imp_ota}, respectively.
Despite requiring fewer IBT resources by ignoring the terms related to $\Y_{b}^{\ulB}$ in \eqref{eq:w_bk_imp_ota}, the combined DL-UL local beamforming design, referred to as \textit{comb. local}, leads to performance degradation by neglecting the construction of $\xib_{b,k}$ in \eqref{eq:w_bk} using $\Y_{b}^{\ulB}$. The separate IBT for DL and UL UEs with and without UL-2 ($\Y_{b}^{\ulB}$) signaling is referred to as \textit{sep. OTA} and \textit{sep. local}, respectively, and they exhibit inferior performance due to increased use of IBT resources. However, the IBT resources for these methods are independent of $|\setK^{\dl} \cap \setK^{\ul}|$. The beamforming design with global CSI at the CU, referred to as \textit{centralized}, assumes a backhaul delay of $5$~ms for both the DL and UL beamformers, and it also requires UE antenna-specific pilot resources in the UL and UE-specific pilot resources in the DL, resulting in worse performance than the IBT methods. However, the performance of the \textit{centralized} design can be improved by optimizing the use of backhaul resources and CSI prediction techniques in temporally correlated channel scenarios.

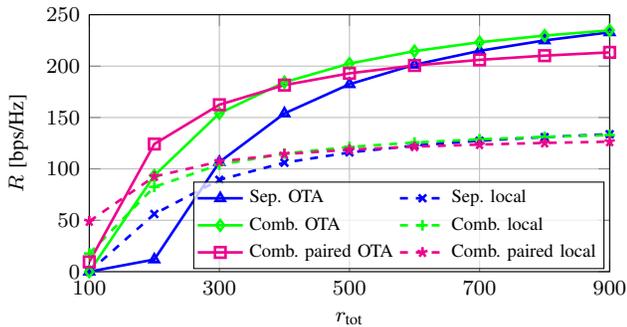
\begin{figure}[t!]
\begin{center}
\input{figures/fig2.tex}
\vspace{-2mm}
\caption{Effective DL-UL sum rate vs. resource block size.}
\label{fig:rateVsrc}
\end{center}
\vspace{-4mm}
\end{figure}

Fig.~\ref{fig:rateVsrc} plots the effective DL-UL sum rate against the resource block size for $t=10$, $|\setK|=44$, $|\setK^{\dl}|=|\setK^{\ul}|=32$, and $|\setK^{\dl\textnormal{-}\ul}|=20$. The proposed combined DL-UL distributed beamforming design with UE pairing is referred to as \textit{comb. paired OTA}. The case where $\Y_{b}^{\ulB}$ is ignored in \eqref{eq:gd_comp} 
is referred to as \textit{comb. paired local}. These designs require fewer IBT resources than \textit{comb. OTA} and \textit{comb. local}, resulting in superior performance for $r_{\textrm{tot}} \le 350$. However, for $r_{\textit{tot}} \le 150$, \textit{comb. paired local} outperforms \textit{comb. paired OTA} due to lower IBT resource usage. For $r_{\textrm{tot}} \ge 350$, the performance of \textit{comb. OTA} surpasses the other methods by leveraging the UE-specific precoder design and moderate IBT resource usage.

\begin{figure}[t!]
\begin{center}
\input{figures/fig3.tex}
\vspace{-2mm}
\caption{{Effective DL-UL sum rate vs. fraction of UEs in the DL~and~UL.}}
\label{fig:rateVsOR}
\end{center}
\vspace{-4mm}
\end{figure}
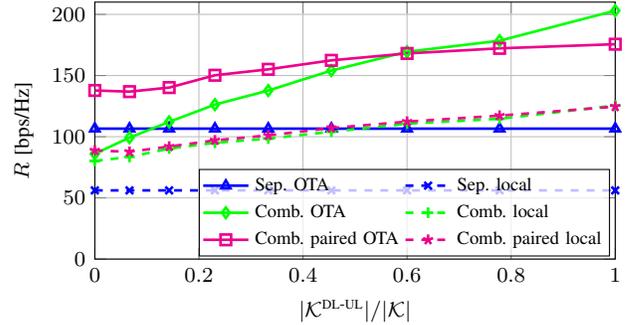

Fig.~\ref{fig:rateVsOR} illustrates the effective DL-UL sum rate against the fraction of UEs in both the DL and UL for $|\setK^{\dl}| = |\setK^{\ul}| = 32$, $t=10$ and $r_{\textrm{tot}}=300$. The number of UEs, $|\setK|$, varies from $64$ to $32$, representing $|\setK^{\dl\textnormal{-}\ul}|/|\setK|$ ratios of $0$ and $1$, respectively. For small value of $|\setK^{\dl\textnormal{-}\ul}|/|\setK|$, the IBT overheads of \textit{comb. paired} OTA and \textit{comb. paired} local are much lower than \textit{comb. OTA} and \textit{comb. local}, resulting in higher performance gains. If the fraction is one, the overhead is the same for all the combined DL-UL methods. However, \textit{comb. OTA} converges faster than the gradient-based \textit{comb. paired} OTA, leading to somewhat better performance. The separate DL and UL IBT methods incur high IBT overhead, thus their performance is inferior to the proposed combined DL-UL distributed beamforming designs for any fraction of DL and UL UEs.

\bibliographystyle{IEEEtran}
\bibliography{IEEEabbr,refs}


\end{document}

%% file: Definitions.tex
\renewcommand{\d}{\mathbf{d}}

\newcommand{\f}{\mathbf{f}}
\newcommand{\g}{\mathbf{g}}
\newcommand{\h}{\mathbf{h}}

\newcommand{\m}{\mathbf{m}}

\newcommand{\p}{\mathbf{p}}

\renewcommand{\v}{\mathbf{v}}
\newcommand{\w}{\mathbf{w}}

\newcommand{\y}{\mathbf{y}}
\newcommand{\z}{\mathbf{z}}

\newcommand{\0}{\mathbf{0}}

\renewcommand{\H}{\mathbf{H}}
\newcommand{\I}{\mathbf{I}}

\newcommand{\X}{\mathbf{X}}
\newcommand{\Y}{\mathbf{Y}}
\newcommand{\Z}{\mathbf{Z}}

\newcommand{\xib}{\boldsymbol{\xi}}

\newcommand{\Phib}{\mathbf{\Phi}}

\newcommand{\setB}{\mathcal{B}}
\newcommand{\setC}{\mathcal{C}}

\newcommand{\setG}{\mathcal{G}}

\newcommand{\setK}{\mathcal{K}}

\newcommand{\setN}{\mathcal{N}}

\newcommand{\Compl}{\mbox{$\mathbb{C}$}}

\newcommand{\Exp}{\mathbb{E}}
\newcommand{\herm}{\mathrm{H}}



%% file: figures/fig1.tex
\begin{tikzpicture}

\begin{axis}[
	width=8.5cm,
	height=5cm,
	xmin=1, xmax=10,
	ymin=0, ymax=210,
    xlabel={$t$},
    ylabel={$R$ [bps/Hz]},
	xtick={1,2,3,4,5,6,7,8,9,10,15,20},
    xlabel near ticks,
	ylabel near ticks,
    x label style={font=\footnotesize},
	y label style={font=\footnotesize},
    ticklabel style={font=\footnotesize},
    legend pos=south east,
    legend cell align=left,
    legend columns=2,
    legend style={at={(0.99,0.01)}, anchor=south east},
	legend style={font=\scriptsize, inner sep=1pt, fill opacity=0.75, draw opacity=1, text opacity=1},
	grid=both,
]

\addplot[line width=1pt, black, dotted, mark=o, mark options={solid}]
table[x=slot, y expr={0.5*\thisrow{Cen}}, col sep=comma] 
{figures/fig1_data/data.txt};
\addlegendentry{{Centralized}};



\addplot[line width=1pt, blue, mark=triangle, mark options={solid}]
table[x=slot, y expr={0.5*\thisrow{SepDLOTA}} , col sep=comma] 
{figures/fig1_data/data.txt};
\addlegendentry{{Sep. OTA}};

\addplot[line width=1pt,  blue, dashed, mark=x, mark options={solid}]
table[x=slot, y expr={0.5*\thisrow{SepDLLocal}} , col sep=comma] 
{figures/fig1_data/data.txt};
\addlegendentry{{Sep. local}};

\addplot[line width=1pt, green, mark=diamond, mark options={solid}]
table[x=slot, y expr={0.5*\thisrow{CombDLOTA}} , col sep=comma] 
{figures/fig1_data/data.txt};
\addlegendentry{{Comb. OTA}};

\addplot[line width=1pt, green, dashed, mark=+, mark options={solid}]
table[x=slot, y expr={0.5*\thisrow{CombDLLocal}} , col sep=comma] 
{figures/fig1_data/data.txt};
\addlegendentry{{Comb. local}};

\end{axis}

\end{tikzpicture}

%% file: figures/fig2.tex
\begin{tikzpicture}

\begin{axis}[
	width=8.5cm,
	height=5cm,
	xmin=100, xmax=900,
	ymin=0, ymax=250,
    xlabel={$r_{\textrm{tot}}$},
    ylabel={$R$ [bps/Hz]},
	xtick={100,300,500,700,900},
    ytick={0,50,100,150,200,250},
    xlabel near ticks,
	ylabel near ticks,
    x label style={font=\footnotesize},
	y label style={font=\footnotesize},
    ticklabel style={font=\footnotesize},
    legend pos=south east,
    legend cell align=left,
    legend columns=2,
    legend style={at={(0.99,0.01)}, anchor=south east},
	legend style={font=\scriptsize, inner sep=1pt, fill opacity=0.75, draw opacity=1, text opacity=1},
	grid=both,
]

\addplot[line width=1pt, blue, mark=triangle, mark options={solid}]
table[x=res, y expr={0.5*\thisrow{SepDLOTA}} , col sep=comma] 
{figures/fig2_data/data.txt};
\addlegendentry{{Sep. OTA}};

\addplot[line width=1pt, blue, dashed, mark=x, mark options={solid}]
table[x=res, y expr={0.5*\thisrow{SepDLLocal}} , col sep=comma] 
{figures/fig2_data/data.txt};
\addlegendentry{{Sep. local}};

\addplot[line width=1pt, green, mark=diamond, mark options={solid}]
table[x=res, y expr={0.5*\thisrow{CombDLOTA}} , col sep=comma] 
{figures/fig2_data/data.txt};
\addlegendentry{{Comb. OTA}};

\addplot[line width=1pt, green, dashed, mark=+, mark options={solid}]
table[x=res, y expr={0.5*\thisrow{CombDLLocal}} , col sep=comma] 
{figures/fig2_data/data.txt};
\addlegendentry{{Comb. local}};

\addplot[line width=1pt, magenta, mark=square, mark options={solid}]
table[x=res, y expr={0.5*\thisrow{CombGrpOTA}} , col sep=comma] 
{figures/fig2_data/data.txt};
\addlegendentry{{Comb. paired OTA}};

\addplot[line width=1pt, magenta, dashed, mark=star, mark options={solid}]
table[x=res, y expr={0.5*\thisrow{CombGrpLocal}} , col sep=comma] 
{figures/fig2_data/data.txt};
\addlegendentry{{Comb. paired local}};

\end{axis}

\end{tikzpicture}

%% file: figures/fig3.tex
\begin{tikzpicture}

\begin{axis}[
	width=8.5cm,
	height=5cm,
	xmin=0, xmax=1,
	ymin=0, ymax=210,
    xlabel={$|\setK^{\dl\textnormal{-}\ul}|/|\setK|$},
    ylabel={$R$ [bps/Hz]},
    xlabel near ticks,
	ylabel near ticks,
    x label style={font=\footnotesize},
	y label style={font=\footnotesize},
    ticklabel style={font=\footnotesize},
    legend pos=south east,
    legend cell align=left,
    legend columns=2,
    legend style={at={(0.99,0.01)}, anchor=south east},
	legend style={font=\scriptsize, inner sep=1pt, fill opacity=0.75, draw opacity=1, text opacity=1},
	grid=both,
]

\addplot[line width=1pt, blue, mark=triangle, mark options={solid}]
table[x=perOvr, y expr={0.5*\thisrow{SepDLOTA}} , col sep=comma] 
{figures/fig3_data/data.txt};
\addlegendentry{{Sep. OTA}};

\addplot[line width=1pt, blue, dashed, mark=x, mark options={solid}]
table[x=perOvr, y expr={0.5*\thisrow{SepDLLocal}} , col sep=comma] 
{figures/fig3_data/data.txt};
\addlegendentry{{Sep. local}};

\addplot[line width=1pt, green, mark=diamond, mark options={solid}]
table[x=perOvr, y expr={0.5*\thisrow{CombDLOTA}} , col sep=comma] 
{figures/fig3_data/data.txt};
\addlegendentry{{Comb.  OTA}};

\addplot[line width=1pt, green, dashed, mark=+, mark options={solid}]
table[x=perOvr, y expr={0.5*\thisrow{CombDLLocal}} , col sep=comma] 
{figures/fig3_data/data.txt};
\addlegendentry{{Comb. local}};

\addplot[line width=1pt, magenta, mark=square, mark options={solid}]
table[x=perOvr, y expr={0.5*\thisrow{CombGrpDLOTA}} , col sep=comma] 
{figures/fig3_data/data.txt};
\addlegendentry{{Comb. paired OTA}};

\addplot[line width=1pt, magenta, dashed, mark=star, mark options={solid}]
table[x=perOvr, y expr={0.5*\thisrow{CombGrpDLLocal}} , col sep=comma] 
{figures/fig3_data/data.txt};
\addlegendentry{{Comb. paired local}};

\end{axis}

\end{tikzpicture}